\documentclass[reprint,prd,nofootinbib]{revtex4-2}
\usepackage{epsfig,amsmath,amsfonts,amssymb}

\begin{document}

\title{Oscillating scalar field as dark matter through the cosmological epochs}

\author{Vladimir A. Popov}
\email{vladipopov@mail.ru} 
\affiliation{Institute of Physics, Kazan Federal University, Kremlevskaya str.~18, Kazan 420008, Russia}

\begin{abstract}
An oscillating scalar field can behave like cold dark matter in the expanding Universe. In relativistic cosmology, this perspective is primarily reasoned by an evolution of a background scalar field and its subhorizon perturbations in a matter-dominated era. A corresponding relativistic description is usually based on the ansatz that the perturbations oscillate in the same way as the background field and on time averaging to find the oscillation amplitudes. In contrast, the presented approach makes no initial assumptions about the perturbation behavior and does not use time averages. It is shown that although the previous treatments are valid only well inside the Hubble scale, the scalar field reproduces the CDM scenarios on the superhorizon scale and also throughout the evolution of the Universe from the beginning of the oscillations. The misunderstanding about the Jeans length scale in the matter-dominated epoch is also clarified. An analog of the Jeans effect is found in a radiation-dominated era. It can prevent an overproduction of small halos for the scalars with $m\sim 10^{-13}$~eV.
\end{abstract}


\maketitle

\section{Introduction}

Today, there is very little doubt that a significant fraction of the matter in the Universe is a hidden dark matter (DM). Its presence in galaxies is evidenced by observational data on the rotation curves \cite{Simon2005High,deBlok2008HIGH,Oman2015unexpected,Lelli2016SPARC} and gravitational lensing \cite{Massey2010dark,Barnabe2011Two-dimensional,Fedorova2016Gravitational}. The CMB anisotropy data indicate a significant DM contribution to the evolution of the Universe \cite{Das2014Atacama,Planck2020Planck}. For more details and references about the current status of DM, see, e.g., the reviews \cite{BERTONE2005Particle,POPOLO2014NONBARYONIC,Freese2017Status,Cirelli2024Dark}. 

The preferred DM model is a cold DM (CDM) formed by collisionless weakly interacting massive particles (WIMP) being nonrelativistic already from the moment of decoupling from a thermal bath \cite{Bertone2010moment,Roszkowski2018WIMP,Arcadi2018waning}. The CDM is in the best agreement with the most reliable observations, and despite some problems, it is a reference model in modern cosmology and astrophysics. Alternative versions of DM are expected to reproduce the main features and properties of the CDM.

The underlying CDM property for the evolution of the Universe is based on the fact that the CDM consists of non-relativistic particles and, therefore, is pressureless. Another notable result is that the CDM provides a fast growth of spatial inhomogeneities in the matter-dominated (MD) stage, which form modern large-scale structures ~\cite{Mukhanov2005Physical,Gorbunov2011Introduction}. The density contrast $\delta$ for subhorizon perturbations satisfies the equation  
\begin{equation}
\label{eq:feature2} 
\ddot\delta + 2H\dot\delta - 4\pi G \rho\delta =
0\,, 
\end{equation} 
where $H$ is the Hubble expansion rate, $\rho$ is the background CDM energy density, and the dot denotes a derivative with respect to the comoving cosmological time. 
This equation gives a solution growing as the scale factor, $\delta\propto a$.

A DM model, which provides similar properties, is based on a massive (pseudo)scalar field \cite{Gorbunov2011Introduction,Turner1983Coherent,Ratra1991Expressions,HWANG1997Roles,HWANG2009Axion,Park2012Axion,Alcubierre2015Cosmological}. 
This model is primarily motivated by axions and axion-like particles (ALPs) as DM candidates~\cite{Preskill1983Cosmology,Abbott1983A,Stecker1983Axions,Marsh2010Ultralight}.

The scalar field with the quadric potential $V(\psi)=\frac12 m^2\psi^2$ evolves in the expanding Universe with the Hubble rate $H\ll m$ according to 
\begin{equation}
\label{eq:ucosmt}
\psi \approx u(t)\cos mt\,, 
\end{equation} 
where  $u\propto a^{-3/2}$ varies slowly compared with the period of the oscillations \cite{Turner1983Coherent,Ratra1991Expressions}. 
This field is effectively pressureless (corresponding components of the stress-energy tensor oscillate as $\cos 2mt$ and vanish when averaged over the period) while the energy density varies as $a^{-3}$. This behavior is characteristic of the CDM. 
When the scalar field dominates the Universe, it provides the same background evolution as non-relativistic matter.

The small spatial inhomogeneities have been studied in \cite{HWANG2009Axion,Park2012Axion} using an \emph{ansatz} that the scalar field perturbations oscillate at the same frequency as the background field. This ansatz has been applied to the scalar field perturbations when the DM dominates in the Universe. Expressions for perturbation amplitudes are derived by averaging over the period. For the long wavelength perturbations, the density contrast satisfies Eq.~(\ref{eq:feature2}) and grows linearly with the scale factor as it takes place for the CDM. 

The perturbations grow when the wavelength is larger than the Jeans length $\lambda_\text{J}\sim 1/\sqrt{mH}$. That is in agreement with the previous non-relativistic estimations~\cite{Hu2000Fuzzy,Sikivie2009Bose}. However, this value is contested in~\cite{Alcubierre2015Cosmological}, where scalar field perturbations were analyzed without time averaging, and the Jeans length was found to be of the order of the Compton wavelength $m^{-1}$. This controversy is resolved in the present paper when discussing subhorizon perturbations.

Some issues also remained beyond the scope of the mentioned studies. Above all, superhorizon perturbations are not covered. Although the perturbations are long wavelengths on the Hubble scale, the corresponding density fluctuations are not expected to grow when considered in the Newtonian gauge. Superhorizon modes are certainly not as significant in galaxy formation as subhorizon modes, but their behavior adds to the overall picture of large-scale structure formation.

Besides, the growing modes in the range $\lambda_\text{J}\lesssim\lambda\lesssim H^{-1}$ enter the horizon at the early radiation-dominated (RD) universe, and the evolution of the field perturbations at this stage need to be analyzed. Clearly, because the Hubble parameter is quite large at the RD epoch, the approximate solution (\ref{eq:ucosmt}) is inapplicable to the ultralight fields while it remains valid for QCD axions with mass $m\sim 10^{-6}-10^{-4}$~eV.

Nevertheless, the results described above give reasons to consider the oscillating scalar field instead of the ordinary CDM. There is a large number of related studies discussing the scalar field as a DM candidate in various contexts~\cite{UrenaLopez2014Nonrelativistic,Schive2014Cosmic,Hui2017Ultralight,Harko2019Jeans,Chavanis2021Jeans,Hartman2022Cosmological} (see also numerous references therein). However, an outstanding question is whether the oscillating scalar field behaves like the CDM at the Hubble scales and beyond the DM-dominated epoch.

The oscillating scalar field is considered here using an asymptotic expansion based on the fast to slow time scales ratio~\cite{Nayfeh2004Perturbation,Kevorkian1996Multiple}. This method provides an appropriate way to separate inhomogeneities according to the spatial scales, thus providing opportunities for a more detailed description of the scalar field perturbations. Similar to~\cite{Alcubierre2015Cosmological}, equations are obtained without averaging techniques. The time average is used only for reading results in terms of the standard CDM. Clearly, this approach reproduces the previously obtained results for the background scalar field and its subhorizon perturbations.

The paper is organized as follows. A description of the background scalar field evolution is presented in Sec.~\ref{sec:bg}. Equations for linear perturbations are discussed in Sec.~\ref{sec:pertEqs}. In the next three sections, perturbations of the scalar field are considered for specific epochs and scales. Finally, the results obtained are discussed.

\section{Background evolution}
\label{sec:bg}

We consider DM as a non-interacting massive scalar field minimally coupled to gravity in the spatially flat, homogeneous, and isotropic spacetime.
The scalar field equation reads as 
\begin{equation} 
\label{eq:KGdimless}
\ddot\psi +3\frac{\dot a}{a}\dot\psi + M^2\psi = 0\,,
\end{equation}
while the scale factor $a$ evaluates according to
\begin{equation}
\label{eq:EE00}
\frac{\dot a^2}{a^2} = \rho \,.
\end{equation} 
Eqs.~(\ref{eq:KGdimless}) and (\ref{eq:EE00}) are represented in the dimensionless form, where the normalized quantities are identified by the same symbols as for the original ones, $\psi(m/\rho_\text{s})^{1/2}\to\psi$, $a/a_\text{s}\to a$, $\rho/\rho_\text{s}\to\rho$, and the index s refers to a nominal starting point within the respective epoch. The dot denotes a derivative with respect to the dimensionless comoving cosmological time $H_\text{s}t\to t$, and $H_\text{s} = 8\pi\rho_\text{s}/3$ is the initial value of the Hubble parameter so that $M=m/H_\text{s}\gg 1$ is a natural large parameter in Eq.~(\ref{eq:KGdimless}), which determines a ratio between time scales in the scalar field evolution.

A solution to Eq.~(\ref{eq:KGdimless}) can be obtained as a two-time dependent uniformly valid asymptotic expansion in powers of $1/M$~\cite{Nayfeh2004Perturbation}
\begin{equation}\label{eq:expansionPSIbg}
\psi = \psi_0(t,T)   + \frac1M\psi_1(t,T)  + \frac1{M^2}\psi_2(t,T)  + \cdots\,,
\end{equation}
where the slow time $t$ corresponds to a cosmological time scale while the fast time
\begin{equation}\label{eq:Tbg}
T=M{\cal S}(t)\,,\quad {\cal S}(t) = S_0(t) +  \frac1{M}S_1(t) + \cdots\,.
\end{equation}

This method implies that Eq.~(\ref{eq:KGdimless}) is decomposed into separate parts corresponding to different powers of $M$, and the equations are solved serially in decreasing order of these powers. Substituting (\ref{eq:expansionPSIbg}) and (\ref{eq:Tbg}) into Eq.~(\ref{eq:KGdimless}), we find that all the terms $\psi_b$ satisfy the equation
\begin{equation}\label{eq:method}
\dot S_0^2\,\partial_T^2\psi_b + \psi_b = F_b\,,
\end{equation}
where the right hand side $F_0=0$, and the next $F_b$ are expressed through functions of the previous orders with respect to $\psi_b$. At the same time, $F_b$ must not produce secular terms in the solutions for $\psi_b$ to provide a uniform validity of the expansion. Eq.~ (\ref{eq:method}) implies $\dot S_0 = 1$ and hence $S_0(t)=t$ resulting in harmonic oscillations with a dominant frequency corresponding to the scalar field mass (with agreement with (\ref{eq:ucosmt})).  The left hand sides of the equations are identical for all functions $\psi_b$, and the secular terms are eliminated if coefficients of $\cos\,T$ and  $\sin\,T$ vanish on the right hand side of (\ref{eq:method}). As a result, we obtain equations that determine the slow time dependence of the functions in (\ref{eq:expansionPSIbg}) and (\ref{eq:Tbg}).

Using this technique, we can find
\begin{equation}
\label{eq:SFexpansion}
\psi(t)= u(t)\cos\,T + O\left( \frac1{M^2}\right)\,,
\end{equation} 
where  $u(t)\propto a^{-3/2}(t)$, and the fast time gives an oscillation phase of the dominant term.
Solution details for subdominant terms depend on which entity dominates the epoch under consideration.

The stage of scalar field domination implies 
\begin{equation}
\rho =  \frac12 \left(\frac{\dot\psi^2}{M^2}+\psi^2\right)\,,
\end{equation}
and equations (\ref{eq:KGdimless}) and (\ref{eq:EE00}) are solved simultaneously using an expansion similar to (\ref{eq:SFexpansion}) for the scale factor. Corresponding solutions are given by
\begin{align}
\label{eq:psiMDbg}&  \psi(t)= u(t)\cos T + \frac1{M^2} \frac{3u^2(t)}{16}\cos 2T + \cdots\,,\\
\label{eq:aMDbg}&  a(t)= a_0(t) + \frac1{M^2} \frac{3u^2(t)}{16}\cos 2T +  \cdots\,,
\end{align}
where the leading term in (\ref{eq:aMDbg}) corresponds to the scale factor driven by non-relativistic matter, 
$a_0(t)\propto t^{2/3}$, 
and the subleading term in (\ref{eq:Tbg}) varies as $S_1(t)\propto u(t)$.

In the RD universe, the energy density $\rho$ on the right hand side of Eq.~(\ref{eq:EE00}) complies with the equation of state $p=\rho/3$ and varies as $\rho= a^{-4}$. The subleading terms can only appear in the third order in the expansion (\ref{eq:SFexpansion}), while $S_1(t)\propto a^{-2}$. 

The current era of an accelerated expansion of the Universe is efficiently simulated by the cosmological constant (referred to as $\Lambda$-dominated era, or $\Lambda$D for short). A major contribution to the background evolution is provided by the constant energy density so that $H=H_\text{s}=$const, and the scale factor $a=e^{t}$ in the dimensionless representation.
For this stage, the solution (\ref{eq:SFexpansion}) reduces to the first term describing single-mode oscillations with a constant frequency, which differs from $M$ by a value of order $1/M$.

The subleading corrections in (\ref{eq:Tbg}) turn out to be insignificant for spatial inhomogeneities and can therefore be dropped from further consideration, while the lower-order terms in (\ref{eq:psiMDbg}) and (\ref{eq:aMDbg}) can be valuable for the compatibility of the perturbed equations.

\section{Equations for linear perturbations}
\label{sec:pertEqs}

In this section we consider equations for small inhomogeneous perturbations of the scalar field in the expanding Universe.
We use the Newtonian gauge for general scalar metric perturbations so that the line element has the form
\begin{equation}
\label{eq:NewtonGuage}
{\,{\rm{d}}} s^2=(1+2\Phi){\,{\rm{d}}} t^2 - a^2(t)(1-2\Psi)\delta_{ij}{\,{\rm{d}}} x^i {\,{\rm{d}}} x^j\,,
\end{equation}
where the spatial coordinates $x^i\ (i=1,2,3)\,$ are reduced to the dimensionless form in the same way as time,  $H_\text{s}x^i\to x^i$.

Isotropic scalar perturbations imply $\Phi=\Psi$. This equality is strictly derived from the perturbed Einstein equations, however it can be exploited from the outset without loss of generality.

The scalar field is now described by the sum
\begin{equation}\label{eq:bg+pert}
\psi(t,\mathbf{x}) = \psi_0(t,T) + \varphi(t,T,\mathbf{x})\,,
\end{equation}
where $\varphi\ll \psi_0$, and the homogeneous background $\psi_0(t,T)$ is represented by (\ref{eq:SFexpansion}). The fast time can now depend on spatial inhomogeneities, 
\begin{equation}\label{eq:fastT}
T=M(t+S(t,\mathbf{x}))\,,
\end{equation}
where a linear perturbation of the phase $S$ is taken into account in the same order as $\varphi$, while the homogeneous corrections of (\ref{eq:Tbg}) are neglected.
Obviously, the representation (\ref{eq:fastT}) can be eliminated by coordinate reparametrizations. That will be briefly discussed in the last section.

The scalar field perturbations obey the equation 
\begin{align}
&\ddot\varphi + 3\frac{\dot a}{a}\dot\varphi - \frac{1}{a^2}\nabla^2\varphi + M^2\varphi
\nonumber\\&
\phantom{\ddot\varphi + 3\frac{\dot a}{a}\dot\varphi}
- M \frac{\partial_T\psi_0}{a^2}\nabla^2 S + 2M^2\partial_T^2\psi_0\dot S
\label{eq:genSFperturb}
\\&
\phantom{\ddot\varphi + 3\frac{\dot a}{a}\dot\varphi}
- 2\left(\ddot\psi_0 + 3\frac{\dot a}{a}\dot\psi_0\right)\Phi - 4\dot\psi_0\dot\Phi
= 0\,,
\nonumber
\end{align}
where $\nabla^2$ is the Laplace operator with respect to the dimensionless coordinates.
The dot is kept to denote the time derivative. It is calculated as $\partial_t + M\partial_T$ for the two-time dependent functions and reduced to the derivative with respect to $t$ for quantities depending only on the slow cosmological time.

Since Eq.~(\ref{eq:genSFperturb}) includes the rapidly oscillating background functions, we also search solutions 
for perturbations as an asymptotic expansion in powers of $1/M$:
\begin{equation}\label{eq:expansionSFpert}
\varphi
= \varphi_0(t,T,\mathbf{x}) + \frac1M\, \varphi_1(t,T,\mathbf{x}) + \frac1{M^2}\, \varphi_2(t,T,\mathbf{x}) + \cdots\,.
\end{equation}

The linearized Einstein equations in the metric (\ref{eq:NewtonGuage}) are
\begin{align}
\label{eq:perturbEE00}
& \frac{1}{a^2}\nabla^2\Phi-3\frac{\dot a}{a}\left(\dot\Phi+\frac{\dot a}{a}\Phi\right) = \frac32\, \delta\rho 
\,,\\
\label{eq:perturbEEii}
& 
\ddot\Phi + 4\frac{\dot a}{a}\Phi 
 + \left(\frac{\dot a^2}{a^2}+2\frac{\ddot a}{a}\right)\Phi 
 = \frac32\, \delta p
\,,\\
\label{eq:perturbEE0i}
& 
\dot\Phi+\frac{\dot a}{a}\Phi
= \frac32\, \delta\mu
\,,
\end{align}
where the right hand sides are produced by the energy-momentum tensor and interpreted as usual: $\delta\rho = \delta T_0^0$ and $\delta p = \delta T_i^i$ (no summation) are normalized energy density and pressure fluctuations respectively, and $\delta\mu$ is a momentum potential, so that  $-\partial_i\delta\mu = \delta T^0_i$. 

In the MD universe these quantities are associated with the scalar field:
\begin{align}
\label{eq:flucdensSF}
& \delta\rho  =
\psi_0\varphi + \frac{1}{M^2}\dot\psi_0\dot\varphi -  \frac{1}{M^2}\dot\psi_0^2\Phi + \frac{1}{M}\dot\psi_0\partial_T\psi_0\dot S 
\,,\\
\label{eq:flucpresSF}
& \delta p =
-\psi_0\varphi +\frac{1}{M^2}\dot\psi_0\dot\varphi -  \frac{1}{M^2}\dot\psi_0^2\Phi + \frac{1}{M}\dot\psi_0\partial_T\psi_0\dot S 
\,,\\
\label{eq:potmomSF}
& \delta\mu =
\frac{1}{M^2}\dot\psi_0\varphi + \frac{1}{M}\dot\psi_0\partial_T\psi_0 S 
\,.
\end{align}
The gravitational potential is expanded in a similar manner to (\ref{eq:expansionSFpert}), $\Phi = \Phi_0 + M^{-1}\Phi_1 + \cdots$, and Eqs.~(\ref{eq:genSFperturb}), (\ref{eq:perturbEE00})--(\ref{eq:perturbEE0i}) are solved simultaneously.

Expressions (\ref{eq:flucdensSF})--(\ref{eq:potmomSF}) remain valid for the energy-momentum tensor of the scalar field over all epochs. 
However, in the RD universe the scalar field has a negligible contribution to the right hand side of (\ref{eq:perturbEE00})--(\ref{eq:perturbEE0i}), and these equations are solved implying the relation $\delta p=\delta\rho/3$. The result is well known~\cite{Mukhanov2005Physical,Gorbunov2011Introduction}; the corresponding gravitational potential is substituted into Eq.~(\ref{eq:expansionSFpert}) as a given function of the slow cosmological time $t$. 

The two time scales give rise to a hierarchy in the wavelength scales of the spatial inhomogeneities. Gradient terms corresponding to long wavelength perturbations are smaller in magnitude and should be taken into account in the lower orders of $1/M$.
To properly include perturbations with different wavelength scales in the expansion (\ref{eq:expansionSFpert}), we introduce a parameter $Q$ identifying inhomogeneity wavelengths with respect to the Hubble scale so that the perturbed functions are taken to depend on the rescaled coordinate $X^i=Q\,x^i$.
Usually, one takes $Q=M^p$ and matches $p$ to get a correspondence with powers of $M$ in an equation solved~\cite{Kevorkian1996Multiple}.  
Appropriate values for Eq.~(\ref{eq:genSFperturb}) are $p=0,\ 1/2,\ 1$.

As we will see in the next section, superhorizon perturbations, as well as perturbations entering the Hubble scale, are described by $Q=1$. These modes are related to cosmic structure formation in both the RD and MD epochs. The value of $Q= M^{1/2}$ 
results in the same equation for the CDM density contrast as in~\cite{HWANG2009Axion}. 
Perturbations of this kind are most responsible for the galaxy formation.
When $Q=M$, the equations describe perturbations of the Compton wavelength scale, which are deep below the Jeans length and thus are out of interest. 

The results obtained in this way are certainly gauge dependent. 
This description is only meaningful in terms of the Newtonian gauge, and it is the gauge that provides an opportunity to compare the results of this work with those given before, as well as with the solutions for the standard CDM.

\section{Perturbations in a matter-dominated universe}

\subsection{Subhorizon scales}
\label{sec:subMD}

The spatial inhomogeneities well inside the Hubble sphere can be considered using the rescaling factor $Q= M^{1/2}$. This value corresponds to subhorizon perturbations with wavelengths $\lambda\gtrsim 1/\sqrt{Hm}$.

Substituting these expansions into the perturbed Einstein equations (\ref{eq:perturbEE00}) and (\ref{eq:perturbEEii}), we find that the equations in the leading orders, $M^2$ and $M$, are satisfied only when $\Phi_{0}$ and $\Phi_{1}$ are independent of the fast time $T$. In addition, $\Phi_{0}$ is independent of the spatial coordinates and, therefore, can depend on the slow time $t$ only. That gives a reason to drop $\Phi_{0}$ from the equations since it does not contribute to the inhomogeneous distribution\footnote{This conclusion can be directly derived from the equations. However, the heuristic reasoning is less tedious.}.

When $\Phi_{0}=0$ the scalar field equation (\ref{eq:genSFperturb}) gives in zero order of $M$
\begin{equation}
\label{eq:KG-10sub}
\partial^2_T \varphi_{0} + \varphi_{0} +2 \dot S \, \partial_T^2\psi_0 -  \frac{\nabla^2 S }{a^2}\,\partial_T\psi_0 = 0\,,
\end{equation}
where nabla now stands for $\nabla=\partial/\partial\mathbf{X}$.
To eliminate the secular terms in the solution, we have to establish $S=0$. As a result, the solution to Eq.~(\ref{eq:KG-10r}) can be represented as
\begin{equation}\label{eq:psi_10}
\varphi_{0}(t,T,\mathbf{X}) = u(t)\left[v(t,\mathbf{X})\cos T + w(t,\mathbf{X})\sin T\right]\,.
\end{equation}

In the next order, the scalar field equation has the form 
\begin{align}
\partial_T^2\varphi_{1} + \varphi_{1} = 
&
-\frac{3\dot a }{a}\partial_t\varphi_{0} - 2\partial_{T}\partial_t\varphi_{0} 
\nonumber\\&
+\frac{1}{a^2}\nabla^2\varphi_{0}
+2 \partial_T^2\psi_0\,\Phi_{1} 
\,.
\end{align}
All of the components on the right hand side produce secular terms and thus must vanish, resulting in
\begin{align}
\label{eq:v10}
& \dot v + \frac{1}{2a^2}\nabla^2 w = 0\,,\\
\label{eq:w10}
& \dot w - \frac{1}{2a^2}\nabla^2 v + \Phi_{1} = 0\,.
\end{align}

In zero order of $M$ we obtain from (\ref{eq:perturbEE00}) and (\ref{eq:perturbEEii})
\begin{align}\label{eq:EE00pert0}
&\frac{1}{a^2}\nabla^2 \Phi_{1} = \frac32\, u^2v\,,\\
\label{eq:EEiipert0}
&\partial_T^2\Phi_{2} = - \frac32\, u^2\left[v\cos 2T + w\sin 2T\right]\,.
\end{align}
Eqs.~(\ref{eq:v10}), (\ref{eq:w10}) and (\ref{eq:EE00pert0}) are reduced to an equation for the density contrast $\delta=2v$. For a plane wave perturbation with wavenumber $K$, this equation reads as
\begin{equation}\label{eq:dc1}
\ddot \delta + 2\frac{\dot a}{a}\dot \delta +\left( \frac{K^4}{4a^4} - \frac34 u^2 \right) \delta = 0\,.
\end{equation}
The same equation was obtained in~\cite{HWANG2009Axion} using the \emph{ansatz} that perturbations behave according to (\ref{eq:psi_10}).

For $K\to 0$, Eq.~(\ref{eq:dc1}) gives a growing solution $\delta\propto a$, which joins with the solution (\ref{eq:vHubblescale}). 
However, contrary to what is argued in~\cite{HWANG2009Axion}, the equation cannot be applied at all scales. Eqs.~(\ref{eq:dotS}) and (\ref{eq:MDhor_v10}) are more relevant to describe perturbations outside and near the horizon.

Perturbations with large wavenumbers do not grow but oscillate. The dividing point between these behaviors is the Jeans wavenumber $k_\text{J}$ when the expression in the parentheses is zero. In dimensional terms, this corresponds to
\begin{equation} \label{eq:kJeans}
k_\text{J}=\left(16\pi G\rho m^2a^4\right)^{1/4}\,.
\end{equation}

The physical Jeans length, $\lambda_\text{J}\propto1/\sqrt{mH}$, is the same as in~\cite{HWANG2009Axion}, although this result was obtained here without time averaging. In contrast, the Jeans length found in~\cite{Alcubierre2015Cosmological} is of the order of the Compton wavelength, $\sim m^{-1}$. This disagreement may be due to the fact that the same series expansion is employed in~\cite{Alcubierre2015Cosmological} on all subhorizon scales including the range $k\gtrsim m$.  In this case, however, the frequency of the fast oscillations for the perturbations becomes dependent on the wavenumbers, and the separation of the fast oscillations from the slowly varying inhomogeneous amplitudes loses its meaning. This point can be understood qualitatively from Eq.~(\ref{eq:dc1}), which gives the frequency of the ``slow'' amplitude oscillations comparable to the frequency of the fast oscillations for $k\sim m$~\footnote{
	A more detailed consideration can be derived from Eq.~(\ref{eq:genSFperturb}) with the rescaling factor $Q=M$. In this case, the fast time for the scalar field perturbations is different from the background one and, moreover, depends on the wavenumber. As a result, the energy density contrast obtained according to (\ref{eq:flucdensSF}) is a rapidly oscillating quantity on the Compton scale.
}.

The Jeans wavenumber could also be obtained by considering the pressure fluctuation analog (\ref{eq:flucpresSF}). 
At the discussed scale, it has the form
\begin{equation}
\delta p = - u^2\left[\tilde{v}\cos 2T + \tilde{w}\sin 2T\right] + \frac1M\,\frac{K^2u^2v}{4a^2}\,,
\end{equation}
where $\tilde{v}$ and $\tilde{w}$ include corrections of $M^{-1}$ order to $v$ and $w$ respectively.
Ignoring the oscillating terms, the effective speed of sound in physical units is
\begin{equation}
c_\text{s}^2=\frac{\delta p}{\delta \rho} = \frac{k^2}{4m^2a^2}\,,
\end{equation}
and the Jeans scale is defined as usual by
\begin{equation}
c_\text{s}^2\,\frac{k^2}{a^2}=4\pi G\rho\,,
\end{equation}
giving (\ref{eq:kJeans}).

The Jeans scale constrains the size of cosmic structures. Eq.~(\ref{eq:dc1}) describes growing perturbations when $K\lesssim 1$, which corresponds to the wavelength $\lambda\gtrsim (mH_0)^{-1/2}$. 
Extending this length to the observed galactic halos, we obtain constraints on the scalar field mass $m\gtrsim 10^{-22}$~eV. 
These ultralight fields are extensively discussed as DM candidates capable of solving the core-cusp and the missing satellites problems without disturbing the large-scale hierarchy~\cite{Hu2000Fuzzy,Schive2014Cosmic,Lee2010Minimum,Lora2012mass,Alexandre2015Dynamical,Fan2016Ultralight,Eby2018Stability}.

\subsection{Hubble scales}

To consider perturbations with wavelengths close to the horizon scale, we use the asymptotic expansions for the field and metric fluctuations in the form (\ref{eq:expansionSFpert}) with the rescaling factor $Q= 1$.  Expanding equations (\ref{eq:perturbEE00})--(\ref{eq:perturbEE0i}) in powers of $M$ similar to the previous case, we find again from the leading orders that $\Phi_{0}$ and $\Phi_{1}$ are independent of the fast time $T$.  However, the next order equations are different from (\ref{eq:KG-10sub}) and (\ref{eq:EE00pert0})--(\ref{eq:EEiipert0}).

The scalar field equation (\ref{eq:genSFperturb}) in zero order of $M$ yields
\begin{equation}
\label{eq:KG-10r}
\partial^2_T \varphi_{0} + \varphi_{0} + 2 \left(\dot S - \Phi_{0}\right) \, \partial_T^2\psi_0  = 0\,.
\end{equation}
The secular terms are eliminated by
\begin{equation}\label{eq:dotS}
\dot S = \Phi_{0}\,,
\end{equation} 
and $\varphi_{0}$ can be represented by (\ref{eq:psi_10}).

In the next order, Eq.~(\ref{eq:genSFperturb}) produces 
\begin{align}
&\partial_T^2\varphi_{1} + \varphi_{1}
+\frac{3\dot a }{a}\partial_t\varphi_{0} + 2\partial_T\partial_t\varphi_{0} 
- \partial_T\psi_{0}\frac{\nabla^2S}{a^2}
\nonumber
\\&\phantom{\partial_T^2\varphi_{1} + \varphi_{1}}
-\left(\frac{3\dot a }{a}\partial_t\psi_{0} + 2\partial_T\partial_t\psi_{0}\right)\Phi_0
\\&\phantom{\partial_T^2\varphi_{1} + \varphi_{1}}
-3\partial_T\psi_{0}\dot\Phi_0 - 2\partial_T^2\psi_{0}\Phi_1
= 0 \,.
\nonumber
\end{align}
Excluding secular terms, we obtain
\begin{align}
\label{eq:MDhor_v10}
&
2\dot v - 3\dot\Phi_0 - \frac{\nabla^2S}{a^2} = 0\,,
\\
&\dot w + \Phi_1=0\,.
\end{align}

Taking into account (\ref{eq:psi_10}), Eqs.~(\ref{eq:perturbEE00}) and (\ref{eq:perturbEE0i}) in zero order of $M$ read as
\begin{align}
\label{eq:EE00MDhor}
&\frac{1}{a^2}\nabla^2 \Phi_{0}
-3\frac{\dot a}{a}\left(\dot\Phi_{0}+\frac{\dot a}{a}\Phi_{0}\right)
= \frac32\, u^2v\,,\\
\label{eq:EE01MDhor}
&\dot\Phi_0+ \frac{\dot a }{a}\Phi_0=\frac34\,u^2S\,.
\end{align}
Eq.~(\ref{eq:EE01MDhor}) has this form because the oscillating terms cancel out.
Similarly, Eq.~(\ref{eq:perturbEEii}) can be split into two parts. 
The first one,
\begin{equation}
\partial_T^2\Phi_2 + \Phi_0\,\partial_T^2a_ 2 = \frac32\,u^2\left[v\cos 2T + w\sin 2T\right]\,,
\end{equation}
where $a_2$ denotes the second term in (\ref{eq:aMDbg}), describes oscillations of the gravitational potential, while the non-oscillating part gives the equation
\begin{equation}\label{eq:EE11MDhor}
\ddot\Phi_0 + 4\frac{\dot a }{a}\dot\Phi_0  = 0\,.
\end{equation}

Eqs.~(\ref{eq:EE00MDhor}), (\ref{eq:EE01MDhor}), and (\ref{eq:EE11MDhor}) are exactly the same as the equations for the metric perturbations of the non-relativistic perfect fluid in the MD universe, where the DM energy density contrast is $\delta=\delta\rho/\rho=2v$, and $S$ acts as a velocity potential for the perfect fluid CDM. Eqs.~(\ref{eq:dotS}) and (\ref{eq:MDhor_v10}) correspond to the perturbed continuity equation.

The dominant term of the gravitational potential is obtained from (\ref{eq:EE11MDhor}) as
\begin{equation}
\Phi_0 = \phi_0(\mathbf{X}) + \tilde{\phi}_0(\mathbf{X}) a^{-5/2} \,
\end{equation}
with arbitrary functions of the spatial coordinates $\phi_0$ and $\tilde{\phi}_0$.
The density contrast $v$ and the phase shift $S$ are determined from (\ref{eq:EE00MDhor}) and (\ref{eq:EE01MDhor}), whereupon Eqs.~(\ref{eq:dotS}) and (\ref{eq:MDhor_v10}) are also satisfied identically.

The decaying mode $\tilde{\phi}_0$ is used to be neglected throughout the perturbation evolution so that $\Phi_0 = \phi_0(\mathbf{X}) $ is time-independent. For plane wave perturbations, $\phi_0(\mathbf{X})=\phi_0(\mathbf{K})\exp(i\mathbf{KX})$, the density variations take the form
\begin{equation}\label{eq:vHubblescale}
\delta = -2\phi_0 \left(1+\frac{K^2a}{3}\right)\,.
\end{equation}

The same relation takes place for the perfect fluid CDM ~\cite{Gorbunov2011Introduction}.
On the superhorizon scales, $K\to 0$, the density contrast is constant in time, while fluctuations entering the horizon are governed by the second term and grow linearly with the scale factor.

It should be mentioned that the scalar field perturbations well outside the Hubble radius were discussed in \cite{Cembranos2016Cosmological}. 
The constant density contrast, $\delta=-2\phi_0$, was obtained by applying  a different \emph{ansatz} for the field fluctuation behavior on this scale, which in terms of (\ref{eq:bg+pert}) can be represented as $\varphi\propto\dot\psi_0$. The result in~(\ref{eq:vHubblescale}) is derived without preconditions and is more general, providing a relation to the subhorizon perturbations.

Also we find for the phase variation $S = \phi_0 t$, which implies that the oscillation frequency fluctuates according to the perturbations of the gravitational potential. 
The oscillations become incoherent outside the Hubble scale, however the background evolution of the Universe remains homogeneous since the scale factor is driven by the amplitude $u$  only, which is independent of the fast time. 

The pressure fluctuations obtained according to (\ref{eq:flucpresSF}) contain only oscillating terms up to and including the order of $1/M$, thus providing an effective zero value on the slow time scale.

\subsection{Decoherence of the superhorizon modes}

The decoherence rate can be estimated qualitatively by considering the spatial two-point correlation function of the scalar field. Assuming that the metric fluctuations are described as a Gaussian random field, the correlation function can be expressed in terms of the correlations of the gravitational potential:
\begin{equation}\label{2pointCF}
\langle\psi(x)\psi(y)\rangle \propto \exp\left\{ 
-m^2t^2 \left( \langle\Phi(x)^2\rangle - \langle\Phi(x)\Phi(y)\rangle \right)
\right\}\,.
\end{equation}

Eq.~(\ref{2pointCF}) shows that the scalar field is incoherent at points farther apart than the Hubble radius. A decoherence rate depends on the variance of the metric perturbations, and the corresponding decoherence time with respect to the cosmological time scale is estimated as
\begin{equation}
\frac{t_\text{dec}}{t}\sim\frac{H}{m\phi_0}
\end{equation}
For the ultralight fields, the ratio $H/m$ can be comparable to the amplitude of the gravitational potential, and the decoherence develops as the Universe evolves.
For the QCD axions, this ratio is extremely small and provides very short decoherence times so that the oscillations in separated regions of the Hubble size are incoherent whenever in the MD epoch.

It also follows from Eq.~(\ref{2pointCF}) that the scalar field, which is incoherent on large scales, tends to become coherent on smaller scales. An appropriate heuristic treatment is that the scalar field oscillates in different phases in distinct regions of the Universe with extents larger than the current horizon scale. This result can be read as independently expanding homogeneous isotropic sub-universes with slightly different parameters.

This picture is similar to that of the Universe dominated by non-relativistic matter, which consists of a collection of some regions lagging behind and others being ahead of the average evolution of the Universe. Although this description is not gauge invariant, it gives a physically clear and relevant representation with respect to the Newtonian gauge.

Perturbations of the oscillation phase are gauge dependent and can be eliminated by coordinate reparametrizations. 
The synchronous gauge is an appropriate one, although it has residual gauge symmetry and the energy density perturbation is different  from the corresponding gauge invariant variable (while those are equal in the Newtonian gauge)~\cite{Mukhanov2005Physical,Weinberg2008Cosmology}. As a result, the density contrast can grow even for modes larger than the Hubble radius~\cite{Gorbunov2011Introduction}.

In our case, however,  the coordinate transformation $x^\alpha_\text{S}=x^\alpha_\text{N}+\xi^\alpha \ (\alpha=0,1,2,3,4)$ excluding the phase $S$ from the solution obtained in the Newtonian gauge implies that $\xi^0=S$. The density contrast is converted as~\cite{Weinberg2008Cosmology}
\begin{equation}
\delta_\text{S} = \delta_\text{N} -\frac{\dot\rho}{\rho}\,\xi^0 =  \delta_\text{N} +3H S\,,
\end{equation}
where the last term is valid for the MD epoch.
Additional terms, which correspond to a residual gauge symmetry and depend only on the spatial coordinates, can be ignored since they give decaying contributions to $\delta_\text{S}$.
Outside the Hubble scale, according to (\ref{eq:dotS}) and (\ref{eq:vHubblescale}), we obtain $\delta_\text{S} =0$. That produces essentially the same picture as discussed above, but from a different perspective: at distances larger than the Hubble radius, the spatial distribution of the density perturbations is diluted, and only the averaged background evolution is observed.

\section{Perturbations in a radiation-dominated universe}

\subsection{Hubble scales}

The developed procedure can be applied, with some remarks, to scalar field perturbations in the RD universe. Above all, the approximation $m\gg H$ now fails for ultralight fields. However, it remains valid for more massive (pseudo)scalars, and the mass determines when the field begins to oscillate. QCD axions with masses near $10^{-6}$ eV are under the approximation even at the QCD phase transition, while masses close to $10^{-22}$ eV become consistent with the condition at the transition to the MD stage.

The scalar field contribution to the expansion rate of the Universe is negligible, and the normalized energy density in Eq.~(\ref{eq:EE00}) behaves as $\rho=a^{-4}$. The solution to Eq.~(\ref{eq:EE00}) in this case is opportunely represented in terms of the conformal time $\eta = \int a^{-1}{\,{\rm{d}}}t$ instead of the comoving time since the scale factor grows according to $a=\eta\,$.

Metric perturbations, determined by radiation fluctuations, now depend on slow time only. The gravitational potential is found as a solution to Eqs.~(\ref{eq:perturbEE00}) and (\ref{eq:perturbEEii}) with $\delta p = \delta\rho/3$. 
The dominant part $\psi_0$ of the scalar field can contribute into the Einstein equations in the same order as the gravitational potential, however these corrections are only time-dependent and have no effect on evolution of inhomogeneities.
Considering plane wave perturbations, we obtain~\cite{Gorbunov2011Introduction}
\begin{equation}\label{eq:gpRD}
\Phi = -3\,\frac{\phi_0(k)}{\chi^2}\left(\cos \chi - \frac{\sin \chi}{\chi}\right)\,,
\end{equation}
where $\chi=k\eta/\sqrt3$.
This solution is joined with the time-independent solution $\Phi=\phi_0$ well outside the Hubble radius, and the decaying mode is ignored. 
On the other hand the metric perturbations entering the horizon ($\chi>1$) are damping oscillations with frequencies and amplitudes depending on the wavenumber.

Near the horizon 
the corresponding spatial scale is described by $Q=1$ so that $k=K$. This implies that all subdominant terms for the gravitational potential vanish, i.\,e. $\Phi=\Phi_0$, and the equations for scalar field inhomogeneities (\ref{eq:dotS}) and (\ref{eq:MDhor_v10}) remain valid.

According to these equations the gravitational potential produces the energy density and the phase perturbations of the scalar field
\begin{align}
\label{eq:RD_v}
&
v = v_0 + \frac92\phi_0\left(\text{Ci}\,\chi- \log \chi - \frac{\sin \chi}{\chi} - \frac{\cos \chi}{\chi^2} + \frac{\sin \chi}{\chi^3}\right)\,,
\\
\label{eq:RD_S}
&S = \frac{9\phi_0}{K^2}\left(1-\frac{\sin \chi}{\chi}\right)\,.
\end{align}

The long wavelength perturbations, $K\to 0$, behave similarly to those in the MD stage. The metric and the energy density perturbations are time independent. The phase fluctuations are determined through the gravitational potential as $S=\phi_0 t$, implying that the scalar field becomes incoherent at the superhorizon scales. 

For perturbations entering the horizon ($\chi\gg 1$), the phase fluctuations become time-independent and vanish for large wavenumbers, while the density contrast increases logarithmically, providing an early growth of inhomogeneities in the same manner as for non-relativistic matter.

\subsection{Subhorizon scales}
\label{sec:MDsubhorizon}

Modes of the gravitational potential (\ref{eq:gpRD}) entering deep inside the horizon, so that the rescaling factor $Q=M^{1/2}$ is applicable, contribute into equation (15) in the order of $1/M$. That means that the gravitational potential is included in Eqs.~(\ref{eq:v10}) and (\ref{eq:w10}) as
\begin{equation}\label{eq:gpRDsub}
\Phi_1 = -9\,\frac{\phi_0(K)}{K^2 a^2}\left(\cos \chi - \frac{\sin \chi}{\chi}\right)\,,
\end{equation}
where we use the rescaled wavenumber $k=KQ$. 
In fact, $\chi$ varies faster than the cosmological time at this scale but continues to be slower than $T$.

As before, we can derive an equation for the density contrast $\delta=2v$
\begin{equation}\label{eq:dceqRDsub}
\ddot \delta + 2\frac{\dot a}{a}\dot \delta + \frac{K^4}{4a^4}  \delta = -\frac{K^2}{a^2}\Phi_1\,.
\end{equation}
In terms of $\chi$, this equation takes the form
\begin{equation}
\delta_{\chi\chi} + \frac1\chi\,\delta_{\chi} + \frac{K^4}{4\chi^2}\,\delta =
\frac{9\phi_0}{\chi} j_1(\chi)\,,
\end{equation}
where the gravitational potential (\ref{eq:gpRDsub}) is rewritten through the spherical Bessel function of the first order.
The solution is given by
\begin{align}\label{eq:dcsolRDsub}
\delta =\,
& 2v_0\cos\left(\frac{K^2}{2}\log\chi \right) +2w_0\sin\left(\frac{K^2}{2}\log\chi \right)
\nonumber\\&
 -\frac{9\phi_0}{K^2} \int\limits_{\chi_0}^\chi j_1(s) \sin\left( \frac{K^2}{2}\log\frac{\chi}{s} \right){\rm{d}}s\,. 
\end{align}

It is clear from this solution that the density perturbations unsupported by the gravitational potentials can only oscillate with amplitudes depending on an initial distribution of the scalar field fluctuations.

\begin{figure}
	\centering
	\includegraphics[width=.95\linewidth]{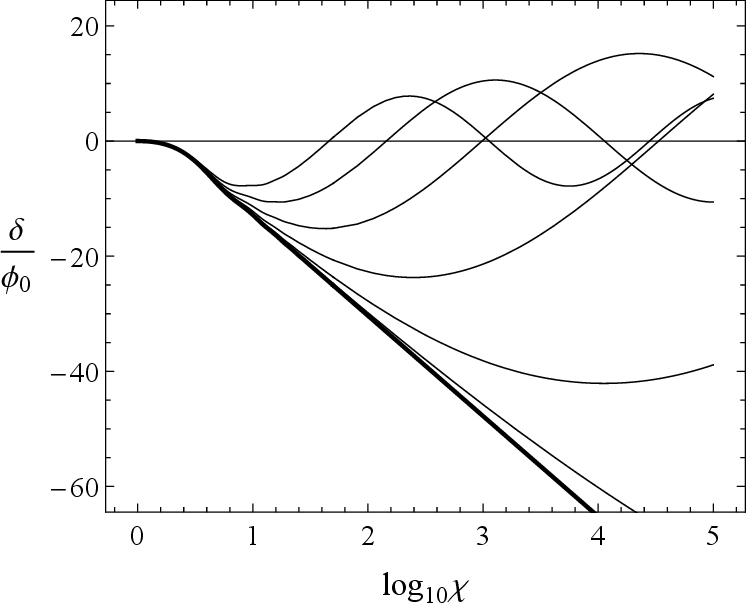}
	\caption{The density contrast of the scalar field produced by the gravitational potential. The curves are calculated according to the last term in (\ref{eq:dcsolRDsub}) with $\chi_0=1$. The thin lines comply with $K=0.4,\ 0.6,\ 0.8,\ 1.0,\ 1.2,\ 1.4$ (from bottom to top).		The thick line corresponds to the solution with $K=0$ represented in Eq.~(\ref{eq:dcKis0RDsub}). }
	\label{fig:fig1}
\end{figure}

The density perturbations driven by the gravitational potentials are described by the last term in (\ref{eq:dcsolRDsub}). The corresponding solutions are shown in Fig.~\ref{fig:fig1}. 
Although these behaviors are forced by an external source, there is an effect similar to the Jeans instability in the MD universe we have concerned in Sec.~\ref{sec:subMD}.

When $K\to 0$, the integral gives
\begin{equation}\label{eq:dcKis0RDsub}
\delta =  9 \phi_0\left( \text{Ci}\chi - \text{Ci}\chi_0 + 
\left(1-\log\frac{\chi}{\chi_0}\right)\frac{\sin\chi_0}{\chi_0} - \frac{\sin\chi}{\chi} \right)\,.
\end{equation}
This solution describes the logarithmic growth of the long wavelength subhorizon perturbations and relates (\ref{eq:dcsolRDsub}) to the perturbations at the Hubble scales represented in (\ref{eq:RD_v}).

When $K$ increases, the scalar field energy density perturbations stop growing and develop into an oscillating regime. However, contrary to the ordinary Jeans effect, there is no definite value of $K$ dividing growing and non-growing modes. Nevertheless, it follows from these solutions that there is an effective short wavelength cutoff for the growing perturbations in the RD epoch. This cutoff can also produce a deficit of dwarf galaxies for the low-mass scalar field DM. 

Indeed, for the ordinary DM, the perturbations in the RD epoch grow logarithmically for all wavenumbers under the horizon. 
This effect is essential for the formation of gravitationally bound structures because the growth in the MD stage alone is not sufficient for the transition to nonlinear evolution~\cite{Gorbunov2011Introduction}. 

In the scalar field model, only the long wavelength subhorizon perturbations are growing in the RD stage. The conditional dividing point between the growing and non-growing modes, $k_\text{d}\sim (mH)^{1/2}\propto  a^{-1}$, evolves in the same way as a physical wavenumber, so that the modes keep their behaviors during this phase. 
When the scalar field becomes a dominant fraction in the Universe, the Hubble rate varies as $a^{-3/2}$ and oscillating modes can turn out above the Jeans radius determined by (\ref{eq:kJeans}).
However, these modes fail 
to reach the nonlinear regime and to form gravitationally bound objects.

The Hubble rate in the RD stage can be represented as
\begin{equation}
H=\left(\frac{g_0}{g} \right)^{1/6}\Omega_\text{rad}^{1/2}(z+1)^2 H_0\,,
\end{equation}
where $H_0$ and $\Omega_\text{rad}$ are the current values of the Hubble parameter and the radiation contribution into the total energy density, $g_0$ and $g$ are the numbers of degrees of freedom at the current and RD epochs respectively. Then, for a current length corresponding to the cutoff scale, we obtain
\begin{equation}
\lambda \sim \frac{2\pi(z+1)}{k_\text{d}} \sim 0.1 \cdot \left(\frac{m}{10^{-13}\ \text{eV}}\right)^{-1/2}\ \text{kpc}\,.
\end{equation}
This scale estimate shows that in the scalar field models with masses close to $10^{-13}$~eV, ordinary galaxies are formed as in the CDM model, while the number of dwarf galaxies is reduced.

The short wavelength cutoff in the MD stage has been considered as a way to solve the missing satellites problem ~\cite{Hu2000Fuzzy,Lee2010Minimum,Lora2012mass}.
According to (4), the Jeans scale corresponding to the small galaxies is characteristic for ultralight fields with $m\sim 10^{-22}$~eV. This effect has stimulated studies of the ultralight bosons as a DM candidate that can combine features of the CDM and warm DM. For more massive scalar particles, the Jeans scale is well below the galactic size, so the cosmic structures grow as in the CDM model. However, for the low-mass scalars, the short wavelength cutoff in the RD stage can lead to a deficit of dwarf galaxies.

\section{Perturbations in a $\Lambda$-dominated universe}

Eqs.~(\ref{eq:perturbEE00})--(\ref{eq:potmomSF}) can also be applied to describe the behavior of the scalar field perturbations in the $\Lambda$D stage. The dominant contribution to the energy density is constant, and the corresponding density perturbation is $\delta\rho_\Lambda=0$. As a result, the metric fluctuations turn out to be related to the scalar field perturbations by the same equations as for the MD universe, but the scale factor is implied to grow exponentially, $a=e^t$. Since the background solution is not provided by the scalar field, the perturbed equations are solved only approximately, neglecting terms that decay rapidly with increasing scale factor~\cite{Gorbunov2011Introduction}.

Density perturbations deep inside the horizon scale are described by Eq.~(\ref{eq:dc1}), where the amplitude of the oscillating scalar field continues to vary as $u=u_0a^{-3/2}$ and the terms in parentheses can be ignored. The equation reduces to
\begin{equation}
\ddot \delta + 2\dot \delta = 0\,,
\end{equation}
which gives the density contrast constant in time if a decaying mode is neglected.

The scalar field perturbations on the Hubble scales are evaluated according to Eqs.~(\ref{eq:EE00MDhor})--(\ref{eq:EE01MDhor}), while Eq.~(\ref{eq:EE11MDhor}) for the gravitational potential is replaced by
\begin{equation}\label{eq:EE11LDhor}
\ddot\Phi_0 +4\dot\Phi_0 +3\Phi_0 = 0\,.
\end{equation}

These equations are similar to the hydrodynamic representation of the CDM and can be examined in the same manner. In the momentum representation, Eq.~(\ref{eq:EE11LDhor}) gives a solution
\begin{equation}
\Phi_0 = \phi_0(K)a^{-1} + \tilde{\phi}_0(K) a^{-3} \,.
\end{equation}
The energy density contrast is found from Eq.~(\ref{eq:EE00MDhor}) as
\begin{equation}\label{eq:dc5}
\delta=2v=-\frac{4}{3u_0^2}[K^2\phi_0(K)-6\tilde{\phi}_0(K)]\,.
\end{equation}

It is clear that the density perturbations do not grow on all relevant scales in the $\Lambda$D stage.

\section{Conclusions}

According to~\cite{HWANG2009Axion,Park2012Axion,Alcubierre2015Cosmological}, a coherently oscillating scalar field has been considered as a cosmologically appropriate entity capable of mimicking features of the CDM. The reasons for that are based on the background behavior of the scalar field energy density decreasing as $a^{-3}$ and on the long wavelength subhorizon density perturbations growing linearly with $a$ in the MD epoch. It is also shown in [2] that the perturbations on the superhorizon scale remain to be constant in this epoch. These results are obtained in frameworks of specific considerations and give separate pieces of a general picture of the Universe with the oscillating scalar field.

The same results have been also obtained in this paper by considering the scalar field through a hierarchy of scales generated by the $H/m$ ratio. This approach naturally establishes a fast time for the oscillations and a slow time for their amplitudes, as well as provides a proper description for perturbations at different spatial scales including decoherence of the scalar field outside the horizon.

As a result, it is clear that there are more overlaps between the oscillating scalar fields and the CDM. The inhomogeneities of the scalar fields well outside the Hubble radius remain constant in time in both the RD and MD stages while the inhomogeneities increase entering the horizon scale. There is a logarithmic growth in the RD stage and a linear one in the MD stage. The inhomogeneities stop growing in the $\Lambda$D stage. These results exactly reproduce the standard CDM scenario.

Besides, oscillations of the scalar field are coherent only inside the Hubble scale. On the superhorizon scales, the oscillation frequency varies according to the spatial distribution of the gravitational potential, and these variations result in decoherence of the oscillations outside the Hubble scale.

The decoherence rate is determined by a parity between the ratio $H/m$ and the amplitude of the gravitational potential.
At the beginning of the oscillations, these  quantities are comparable, and the decoherence proceeds at a rate of the expansion of the Universe. The corresponding stage for the QCD axions is realized in the RD epoch, while the ultralight fields come under these conditions in the MD epoch. When axions dominate in the Universe, the $H/m$ ratio is extremely small. In this case, there is instantaneous decoherence with respect to cosmological time scales.

A valuable virtue of the scalar field as a DM model is that it provides a density cusp avoidance in the halo, as well as a natural mechanism to reduce the abundance of small halos for the particular field masses. The latter property was a motivation to study ultralight bosons with $m\sim 10^{-22}$~eV as DM candidates. In this work, a similar opportunity was revealed for light scalars with $m\sim 10^{-13}$~eV.

It should be noted that recent surveys have increased the number of satellites observed in the Milky Way \cite{Koposov2015Beasts,Drlica-Wagner2015EIGHT,Homma2024Final}. Besides, a number of studies have found that feedback processes and reionization prevent star formation in low-mass halos \cite{Bullock2000Reionization,Brown2014THEQUENCHING}. 
These results count in favor of the fact that the number of satellite galaxies in the Milky Way can be consistent with the number predicted by CDM~\cite{Kim2018Missing,Read2019Abundance}.
In this case, a number of DM candidates are ruled out, including the scalars mentioned above. Nevertheless, even in these circumstances, the oscillating scalar field remains the appropriate DM model in a wide range of masses.

As we can see, the oscillating scalar field model effectively produces the same scenario as the ordinary CDM on all cosmological scales throughout the evolution of the Universe. At the same time, it is possible that in the early Universe, the scenario can be somewhat different due to the interactions between axions and radiation or particles~\cite{Ni2008From,Chelouche2009Cosmic,Stadnik2015Axion,Balakin2015Spin,Balakin2021Interaction}. This problem is going to be studied in a future paper.

\acknowledgments 
The work is carried out in accordance with the Strategic Academic Leadership Program ``Priority 2030'' of the Kazan Federal University.


\bibliography{lit2}

\end{document}